\documentclass[twocolumn,nofootinbib,amsmath,amssymb,prd]{revtex4}
\usepackage{hyperref}
\usepackage{mathrsfs}
\usepackage{latexsym}
\usepackage{amsmath}
\usepackage{amsbsy}
\usepackage{amssymb}
\usepackage{epsfig}
\usepackage{graphicx}
\usepackage{dcolumn}
\usepackage{graphicx}
\usepackage{bm}
\usepackage{natbib}
\usepackage{rotating}
\usepackage{pbox}
\usepackage{mathtools}

\begin{document}

\title{Electrostatic Potential of a Point Charge in a Brans-Dicke Reissner-Nordstr\"{o}m Field}

\author{M. Watanabe	}
	\email{maya.watanabe@monash.edu}, 

\author{A. W. C. Lun}
	\email{anthony.lun@monash.edu}

\affiliation{Monash Centre for Astrophysics\\
		School of Mathematical Sciences, Monash University\\
		Wellington Rd, Melbourne 3800, Australia}
\received{May 28, 2013}

		\begin{abstract}
		We consider the Brans-Dicke Reissner-Nordstr\"{o}m spacetime in isotropic coordinates and the electrostatic field of an electric point charge placed outside its surface of inversion. We treat the static electric point charge as a linear perturbation on the Brans-Dicke Reissner-Nordstr\"{o}m background. We develop a method based upon the Copson method to convert the governing Maxwell equation on the electrostatic potential generated by the static electric point charge into a solvable linear second order ordinary differential equation. We obtain a closed form fundamental solution of the curved space Laplace equation arising from the background metric, which is shown to be regular everywhere except at the point charge and its image point inside the surface of inversion. We also develop a method that demonstrates that the solution does not contain any other charge that may creep into the region that lies beyond the surface of inversion and which is not covered by the isotropic coordinates. The Brans-Dicke Reissner-Nordstr\"{o}m spacetime therefore is linearly stable under electrostatic perturbations. This stability result includes the three degenerate cases of the fundamental solution that correspond to the Brans Type 1, the Reissner-Nordstr\"{o}m and the Schwarzschild background spacetimes. 
		\end{abstract}
		
		\maketitle

\section{INTRODUCTION}

The effect of gravitation on electromagnetic phenomena is of great interest due to its applications in both particle and astrophysics. To the astrophysicist, such study sheds light on phenomena occuring around black holes and other strong gravitational sources. Of particular interest to the authors of this paper are the closed form fundamental solutions for electric potential that can be found for these situations. Such solutions, when they exist, are particularly interesting as they provide the basis for research on, amongst others, particle self-interaction [eg. \cite{linet1984}, \cite{leaute_self-interaction_1985}, \cite{Piazzese1991} \cite{drivas2011}] and electromagnetic phenomena around wormholes [eg. \cite{Beskin2011}, \cite{Khusnutdinov2010}, \cite{Khusnutdinov2005}, \cite{bezerra2009}]. See also \cite{Dehghani2006}, \cite{Hendi2008} and \cite{Hendi2012}. This paper investivates the stability of a class of electrovac Brans-Dicke spacetimes linearly perturbed by a static electric point charge. 

In 1927, Whittaker \citep{Whittaker1927b} investigated electric phenomena in gravitational fields including the study of the electrostatic potential generated by a static electric point charge in a Schwarzschild and a quasi-uniform gravitational background. Using the method of separation of variables, he was able to find an infinite series solution describing the former and a closed form solution for the latter. 

Shortly after in 1928, Copson \cite{Copson1928} used Hadamard's \citep{hadamard1923} theory  of ``elementary solutions" (referred to in recent literature as fundamental solutions) of partial differential equations to not only reproduce Whittaker's original expression for the quasiuniform field but to go on and derive an exact closed form expression for the potential generated by a static electric point charge in a Schwarzschild spacetime written in isotropic coordinates. Copson noted his result differed from Whittaker's infinite series solution by a zeroth order term (see Section \ref{surfintandbc} for further discussion on Whittaker's and Copson's solutions).

Independently in the 1970s, Cohen and Wald \cite{cohen_point_1971} and Hanni and Ruffini \cite{hanni_lines_1973} used the method of separation of variables to express the electrostatic potential generated by a static electric point charge in a Schwarzschild background as an infinite series which concurred with Whittaker's earlier result.

Copson's result was ammended by Linet \cite{Linet1976} who applied the boundary condition at infinity and an asymptotic expansion of Copson's solution to prove that it was for not one but two charges, the second residing within the horizon. Linet resolved this issue by excising the second charge, his result coinciding with those found using multipole expansions by Cohen and Wald \cite{cohen_point_1971}, Hanni and Ruffini \cite{hanni_lines_1973} and Whittaker \cite{whittaker_course_1927}. Linet was also able to transform Copson's fundamental solution from isotropic coordinates into the usual Schwarzschild coordinates. 

Using the Copson-Hadamard method Linet went on to derive expressions for the potential of a static electric point charge in the Reissner-Nordstr\"{o}m field with Leaute \citep{leaute_electrostatics_1976} and in a Brans-Dicke-Schwarzschild field with Teyssandier  \citep{Linet1979}. Copson revisited his method of solution in 1978 \citep{Copson1978}, developing a closed form solution for the potential generated by a static electric point charge (what he terms an ``electron") in a Reissner-Nordstr\"{o}m background field. His result again differed to that obtained by Leaute and Linet \citep{leaute_electrostatics_1976} due to a different choice of boundary conditions, which will be discussed in Section \ref{surfintandbc}.

The Copson method for solving for the electric potential involves identifying a new independent variable that converts the governing Maxwell equation into a linear second order ordinary differential equation. Linet and co-authors adopted the form of Copson's independent variable to consider the Reissner-Nordstr\"{o}m and Brans-Dicke-Schwarzschild cases and were able to find the corresponding governing equations for the above mentioned cases as second order ordinary differential equations (see also \citep{Linet2005}). The results obtained by these authors were on a case-by-case basis. Here we introduce a method by which one is able to extract the form of the new independent variable and obtain the general second order ordinary differential equation for all the electrovac spherically symmetric Brans type solutions which are reducible to the Schwarzschild and Reissner-Nordstr\"{o}m black hole solutions in the Einstein theory.

As Copson's method involves solving a linear second order ordinary differential equation which would naturally produce two linearly independent solutions, it is necessary to impose appropriate boundary conditions that would allow one to determine the relationship between the two constant coefficients of the general solution. As mentioned earlier, it was due to a different choice of boundary condition that Copson's result differed from that found by Linet and co-authors, thus it is clear to see how the interpretation of the solution hinges upon the choice of boundary condition. As a result of his choice of boundary conditions Copson's solution exhibited two charges which was in contradiction to Hadamard's theory of ``elementary" solutions that stipulates that there must exist only one singular point. Here we impose a boundary condition such that a Gauss' Law type integral over any closed surface in space not enclosing the perturbing charge must vanish even if that region contains a surface of inversion which exists in all Brans type I solutions which are reducible to the Schwarzschild and Reissner-Nordstr\"{o}m black hole cases (see Theorem 1 in Section \ref{STFT}). This boundary condition proves to be sufficient in determining the relationship between the constant coefficients such that they are in agreement with known multipole solutions found using the method of separation of variables \citep{whittaker_course_1927}, \citep{cohen_point_1971}, \citep{hanni_lines_1973} and those found using Hadamard's definition of ``elementary" solutions \citep{Linet1976}, \citep{leaute_electrostatics_1976}, \citep{Linet1979}. 

In Section \ref{STFT} we give a detailed overview of the Brans-Dicke Reissner-Nordstr\"{o}m background which is the exact solution for the gravitational field generated by a point charge in a scalar-tensor field. The general Brans-Dicke electrovac solutions has six constants of integration, two of which can be determined by scaling the coordinates $r$ and $t$. Luke and Szamosi \cite{Luke1972} show that the remaining four contants of integration can be constrained such that the solution reduces to the Reissner-Nordstr\"{o}m solution in Einstein's theory. The salient features of these Brans-Dicke metrics is that in isotropic coordinates a surface of inversion separates the solution into two regions. This results in a double-covering of the spacetime region corresponding to the exterior of a non-rotating black hole in General Relativity. It is also important to detail the choice of constants and their subsequent physical interpretations as this will influence our choice of boundary condition (see Appendix A for more detail). As demonstrated by Arnowitt, Deser and Misner \citep{Arnowitt1960}, \citep{Arnowitt1960a}, the interpretations of the source terms of a spherically symmetric spacetime in isotropic coordinates requires careful analysis. The results in this section, together with the analysis in Appendix A, extends some of the results in \citep{Arnowitt1960} and \citep{Arnowitt1960a} done using the ADM technique on Schwarzschild and Reissner-Nordstr\"{o}m spacetimes. We also state the appropriate choice of parameter values for the static spherically symetric Brans-Dicke electrovac solutions as required by the weak-field approximation. We briefly discuss observational constraints on $\omega$ as this is of particular interest due to the fact that scalar fields and the variable cosmological ``constant" have become two of the most popular candidates for dark energy \citep{Bergmann1968}, \citep{Wagoner1970}, \citep{Berman1989}, \citep{Nakamura2006}, \citep{Thushari2010a}.

In Section \ref{EP}, we briefly outline Hadamard's theory of fundamental solutions of curved space Laplace equations containing first order terms. In Section \ref{exten} we extend Copson's method to find the first four terms of the fundamental solution describing the potential generated by an static electric point charge placed outside the surface of inversion in a Brans-Dicke Reissner-Nordstr\"{o}m background. We then develop a method of identifying the new independent variable using the Brans-Dicke field equations as outlined in Appendix A. 

In Section \ref{solving}, we solve the linear second order differential equation to give us a closed-form solution, which can be used to construct the fundamental solution that represents the electric potential generated by a point charge residing outside the surface of inversion in a Brans-Dicke-Reissner-Nordst\"{o}m spacetime. It is important to note that due to the nature of the background metric in isotropic coordinates, the region interior to the surface of inversion is an exact copy of the exterior. Therefore the closed-form solution obtained has an additional singular point at the inversion point of the perturbing static electric point charge. 

In Section \ref{surfintandbc}, we introduce a boundary condition that will allow one to determine the relationship between the two constant coefficients and essentially eliminate the singularity that creeps into the spacetime region that lies beyond the inversion
surface and which is not covered by the isotropic coordinates. Hence we obtain a process to derive the fundamental solution for a class of curved space Laplace equations containing first order terms thus making it unnecessary to compare with multipole expansion solutions. 

Lastly, in Section \ref{specificcases} we show how our method also yields the fundamental solutions of the three known cases (Schwarzschild, Reissner-Nordstr\"{o}m, Brans-Dicke Type I) which are summarized in Table \ref{tab:tab}.

In Appendix A, details of how the scalar field, the metric functions and the electrostatic potential are all essentially a combination of the metric variable $r^2 \phi e^{\alpha + \beta}$ are discussed. Appendix A also outlines how the Brans-Dicke Reissner-Nordstr\"{o}m background metric can be determined. The surface integral inner boundary condition is outlined in Appendix B for the Brans-Dicke Reissner-Nordstr\"{o}m background .

\section{Scalar-Tensor Field Theory}\label{STFT}
The field equations in the Brans-Dicke theory are

\begin{eqnarray} \label{fieldone} \nonumber
R_{a b}-\frac{1}{2}g_{a b}R &=& \frac{8 \pi T_{a b}}{c^{4} \phi} + \frac{1}{\phi} (\nabla_{a} \partial_{b} \phi - g_{a b} \square \phi) 
\\ && + \frac{\omega}{\phi^{2}}(\partial_{a} \phi \partial_{b} \phi - \frac{1}{2}g_{a b} g^{c d} \partial_{c} \phi \partial_{d} \phi) 
\end{eqnarray}
\begin{eqnarray}\label{fieldtwo}
\square \phi= \frac{8 \pi T}{(2 \omega + 3)c^{4}}, 
\end{eqnarray}
where
\begin{eqnarray}
\square \phi &:=& \nabla_{b} (g^{ab} \partial_{a} \phi) = \frac{1}{\sqrt{-g}} \partial_{b} (\sqrt{-g}g^{ab} \partial_{a} \phi)
\end{eqnarray}
and $\square$ is the scalar wave operator. 


Here the notations have their usual meaning. The contribution of the electromagnetic field, encoded in the Faraday tensor $F_{ab}$, to the energy-momentum tensor is
\begin{eqnarray}
T_{ab} = F_{ac}F^{c}_{b} - \frac{1}{4}g_{ab}F_{cd}F^{cd} \; , \: T^{a}_{a} = 0,
\end{eqnarray}
where $F_{ab}$ satisfies the source-free Maxwell equations
\begin{eqnarray} \label{maxwell}
\nabla_b F^{a b}= 0 \: , \: \nabla_{[c} F_{ab]} =0.
\end{eqnarray}

Following the method of Luke and Szamozi \citep{Luke1972} while at the same time conforming to the choice of boundary conditions in Brans \citep{Brans1962} (see equations (\ref{weakfieldone}) to (\ref{weakfieldthree})) one can verify that an electrically charged Brans-Dicke field that reduces to the Reisnner-Nordstr\"{o}m solution in isotropic coordinates when the long-range field equals the reciprocal of the gravitational constant, i.e. $\phi = (G_{0})^{-1}$, can be summarized as follows (see Appendix A for a brief derivation).

\textbf{Theorem 1}
\textit{A static spherically symmetric electrically charged Brans-Dicke-Reissner-Nordstr\"{o}m (BDRN) solution of equations (\ref{fieldone}), (\ref{fieldtwo}) and (\ref{maxwell}) in isotropic coordinates $(t,r,\theta, \phi)$ is given by the line element}
\begin{eqnarray} \label{lineelement}
ds^{2}=-c^{2}e^{2\alpha(r)}dt^{2}+e^{2\beta(r)}[dr^{2}+r^{2}(d\theta^{2}+\sin\ \theta d\phi^{2})],
\end{eqnarray}
\textit{where the static electric potential $V_0(r)$, the Faraday tensor $F_{ab}$ and the corresponding energy-momentum tensor $T^{a}_{b}$ are:}
\begin{eqnarray}
V_0(r) &=& Q \int_{\infty}^{r} \frac{e^{\alpha(r)-\beta(r)}}{r^{2}}dr, \label{V}\\
F_{ab}&=& -cV'_0(r) \begin{pmatrix}
0 & 1 & 0 & 0 \\ 
-1 & 0 & 0 & 0 \\ 
0 & 0 & 0 & 0 \\ 
0 & 0 & 0 & 0
\end{pmatrix}, \label{Fab}\\
T^{a}_{b}&=& -\frac{e^{4\beta(r)}Q^{2}}{2r^{4}} \begin{pmatrix}
1 & 0 & 0 & 0 \\ 
0 & 1 & 0 & 0 \\ 
0 & 0 & -1 & 0 \\ 
0 & 0 & 0 & -1
\end{pmatrix}. \label{Tab}
\end{eqnarray}

\textit{The reduced long-range scalar field wave equation derived from equation (\ref{fieldone}) is }
\begin{eqnarray} \label{scalarfield}
[r^2 \exp(-\alpha(r) + \beta(r)) V'_0(r)]'=0.
\end{eqnarray}

\textit{The metric functions $e^{2 \alpha(r)}$ and $e^{2 \beta(r)}$ are}
\begin{eqnarray} 
e^{2\alpha(r)} &=& \frac{e^{2\alpha_0} \left| \frac{r-B}{r+B} \right|^{\frac{2}{\lambda}}}{\left( p_{+}^{2}-p_{-}^{2}\left| \frac{r-B}{r+B} \right|^{\frac{C+2}{\lambda}} \right)^{2}}, \label{e2alpha}\\ \nonumber
e^{2\beta(r)} &=&e^{2\beta_0}  \left(1 + \frac{B}{r} \right)^4  
\left| \frac{r-B}{r+B}\right| ^{2 \left( \frac{\lambda-C-1}{\lambda} \right)} \\&& \times \left( p_{+}^{2}-p_{-}^{2}\left| \frac{r-B}{r+B} \right|^{\frac{C+2}{\lambda}} \right)^{2} , \label{e2beta}
\end{eqnarray} 
\textit{and the long range scalar field $\phi(r)$ is}
\begin{eqnarray}
\phi = \phi_{0} \left| \frac{r-B}{r+B} \right|^{\frac{C}{\lambda}}. \label{phi}
\end{eqnarray}

\noindent \textit{The functions $V_0(r)$, $e^{2\alpha(r)}$, $e^{2\beta(r)}$ and $\phi(r)$ are defined for all non-negative $r$ except at $r=B$. The boundary values $e^{\alpha_0}$ and $e^{\beta_0}$ can be rescaled to unity by scaling the $t$ and $r$ coordinates respectively. The nine parameters $Q, B, p_{+}^{2}, p_{-}^{2}, \lambda, C, \phi_{0}, e^{2\alpha_0}$ and $e^{2\beta_0}$ in equations (\ref{e2alpha}) to (\ref{phi}) are related via }

\begin{equation} \label{lambda}
4\lambda^{2}= (2 \omega +3) C^2 + (C+2)^2, \rule{0.5cm}{0cm} \omega \geq -\frac{3}{2},
\end{equation}
%

\begin{equation}\label{B}
B = \frac{1}{2} \sqrt{m_{B}^{2}-q_{B}^{2}} ,
\end{equation}

\begin{equation} \label{mnq}
m_{B}:= \frac{2M}{c^{2}\phi_{0}} \frac{e^{\beta_0 }\lambda}{C+2}, \rule{0.5cm}{0cm} q_{B}= 2\sqrt{ \frac{4 \pi}{\phi_{0}}} \frac{Q}{c^{2}} \frac{e^{\beta_0 }\lambda}{C+2},
\end{equation}

\begin{equation} 
p_{\pm}^{2} = \frac{m_{B} \pm \sqrt{m_{B}^{2}-q_{B}^{2}} }{2 \sqrt{m_{B}^{2}-q_{B}^{2}}}  \label{pnp},
\end{equation}

\begin{equation} \label{p-}
p_{+}^{2} - p_{-}^{2} =1,
\end{equation}

\noindent \textit{where $M$ and $Q$ are non-negative real constants and are identified, respectively, as mass measured in conventional units (kg) and charge measured in electrostatic units (e.s.u), which has the dimensions of $[mass]^{\frac{1}{2}}[length]^{\frac{3}{2}}[time]^{-1}$. Here $\omega$ is the coupling constant that couples the scalar field to matter, while $c$ is the speed of light in a vacuum.}

Taking into account equations (\ref{lambda}) to (\ref{mnq}), there remain only four essential parameters in the BDRN solution. We adopt the independent parameter set ${M, Q, C, \phi_0}$.
\begin{enumerate}
 \item The choice of the physical parameters of mass, $M$, and charge, $Q$, in the characterization of the BDRN metric is natural.
 \item As opposite charges neutralize one another, in most astrophysical applications it is reasonable to assume $M \geq \sqrt{4 \pi \phi_{0}}Q \geq 0$, and hence the parameter $B$ in equation (\ref{B}) is non-negative.
 \item $\phi_0$ is the value of the long range scalar field at spatial infinity. It has the dimensions of $[mass][length]^{-3}[time]^{2}$.
 \item 
The parameter $C$ is dimensionless and relates to the local strength of the long-range scalar field $\phi(r)$. Equation (\ref{lambda}) gives $\lambda^{2}$ as a quadratic expression in $C$ with the discriminant $\bigtriangleup= -(2\omega +3)$. Thus when $\omega >- \frac{3}{2}$, $C$ is real and $\sqrt{\frac{2\omega+3}{2\omega+4}}<\vert\lambda \vert< \infty$. Constraining the BDRN solutions to conform with the weak-field approximation (see \citep{Brans1961}) we expand the metric functions and the scalar field, equations (\ref{e2alpha}), (\ref{e2beta}) and (\ref{phi}), to the order of $1/r$, and obtain the following restrictions on the parameters:
\begin{eqnarray} \label{weakfieldone}
\alpha_0&=&\beta_0=0,\\
\phi_{0} &=& \frac{1}{G_{0}} \left( \frac{2 \omega+4}{2 \omega +3} \right),\\ \label{weakfieldtwo}
\lambda &\cong& \sqrt{\frac{2\omega+3}{2\omega+4}}\\
C &\cong& -\frac{1}{\omega+2}, \label{C} \\
m_B &\cong& \frac{M}{c^4\phi_0} \sqrt{\frac{2\omega+4}{2\omega+3}}, \\
q_B &\cong& \sqrt{\frac{4\pi}{\phi_0}} \frac{Q}{c^2} \sqrt{\frac{2\omega+4}{2\omega+3}}, \label{weakfieldthree}
\end{eqnarray}
where $G_0$ is defined as the gravitational constant (for the BDRN spacetimes), while $G$ denotes Newton's universal constant of gravity (see Case 2 and Case 3 below).

\item  \label{obs} Observational constraints put even stronger requirements on the values of $\omega$. The latest results obtained from the Cassini-Huygen experiment \citep{Bertotti2003} put the value of $\omega$ at over $40000$. The coupling constant $\omega$ represents the strength of the coupling between the scalar field and the gravitational field. Therefore its value is of great importance in any discussion regarding (a) the existence and properties of Brans-Dicke black holes, and (b) candidates for dark energy. 

\item When an inversion is applied, that is, transforming from $r$ to $r* = \frac{B^{2}}{r}$, the region $B <r < \infty$ is mapped one to one onto the region $0 < r < B$. Under such a reflection at the sphere of $r_{BDRN}=B$, the functions $V_0(r)$, $e^{2\alpha(r)}$, $\phi(r)$ and the line element (\ref{lineelement}) remain invariant while the metric function $e^{2\beta(r)}$ is transformed into $e^{2\beta(r^{*})}=\frac{r^{4}}{B^{4}}e^{2\beta(r)}$ and the flat 3-metric $d\ell^{2}:=[dr^{2}+r^{2}(d\theta^{2} + \sin^{2} \theta d \phi^{2})]$ is mapped conformally onto the flat metric $d(\ell^{*})^{2}=\frac{r^{4}}{B^{4}}d\ell^{2}$. Therefore the spherical surface at $r_{BDRN}:= B =\frac{1}{2}\sqrt{m_{B}^{2}-q_{B}^{2}}$ is an inversion surface in the sense that the BDRN solution in isotropic coordinates is invariant under the geometric inversion transformations $r(r*)=B^{2}$. The two copies of the BDRN spacetime, one exterior to and the other interior to the inversion surface, are identical. At the spherical surface of inversion $r_{BDRN}=B$, the line element (\ref{lineelement}) is singular. It is the inaccessible boundary of the two identical copies of the BDRN spacetime in isotropic coordinates. Throughout this article we use the exterior copy where $B< r< \infty$, unless stated otherwise. This will have important consequences (see below) on how to interpret the Copson-Hadamard method \citep{Copson1928} in the construction of the fundamental solution to the Laplace equation of a perturbed electrostatic potential in a BDRN background solution. 

\item \label{bhadra} An investigation by \citep{Bhadra2005} found that Brans Type I solutions may represent an external gravitational field for nonsingular spherically symmetric matter sources. They concluded however, that Brans-Dicke black holes cannot exist, as a condition equivalent to equation (\ref{C}) (that is, the weak-field approximation) would require that $-2>\omega>-(2+\frac{1}{\sqrt{3}})$; a requirement which clearly violates observational constraints \citep{Bertotti2003}. 
 \end{enumerate}

By choosing various combinations of the four independent parameters $M, Q, C$ and $\phi_0$ to vanish, we obtain the following limiting solutions:

\textbf{Case 1:} Brans Type I (BS) metric in isotropic coordinates

By setting the charge parameter $Q$ to zero, it implies that $B= \frac{m_{B}}{2}= \frac{M}{2c^{2}\phi_{0}}, p_{+}^{2}=1$ and $p_{-}^{2}=0$. We recover the Brans Type I metric \citep{Brans1962} of the Brans-Dicke theory:
\begin{eqnarray}
\phi(r)&=& \phi_{0} \left| \frac{r-B}{r+B} \right|^{\frac{C}{\lambda}},\\
e^{2\alpha(r)} &=& e^{2 \alpha_0} \left|\frac{r-B}{r+B} \right|^{\frac{2}{\lambda}},\\
e^{2\beta(r)} &=&  e^{2 \beta_0} \left(1+\frac{B}{r}\right)^4  \left| \frac{r-B}{r+B} \right|^{2 \left( \frac{\lambda-C-1}{\lambda} \right)},\\
\end{eqnarray}
where $B < r < \infty$ and the inversion spherical surface is at $r_{BS} = B = \frac{M}{2c^{2}\phi_{0}}$ and is a curvature singularity.

\textbf{Case 2:} Reissner-Nordstr\"{o}m (RN) metric in isotropic coordinates

By setting the parameters $C=\alpha_0=\beta_0=0$, it implies that $\lambda^{2}=1, \phi_{0}=(G)^{-1}, B= \frac{1}{2}\sqrt{m^{2}-q^{2}}, p_{+}^{2}= \frac{m+\sqrt{m^{2}-q^{2}}}{2\sqrt{m^{2}-q^{2}}}$ and $p_{-}^{2}= \frac{m-\sqrt{m^{2}-q^{2}}}{2\sqrt{m^{2}-q^{2}}}$, where $m:=\frac{GM}{c^{2}}$ and $q:= \frac{\sqrt{4 \pi G}Q}{c^{2}}$ are respectively the mass and the electric charge measured in grativtational units. The metric functions reduce to the usual Reissner-Nordstr\"{o}m solution in isotropic coordinates:
\begin{eqnarray} \label{RN1}
e^{2\alpha(r)}&=& \frac{\left(r- \frac{\sqrt{m^2-q^2}}{2} \right)^2 \left(r+ \frac{\sqrt{m^2-q^2}}{2}\right)^2}{ \left(r+\frac{m-q}{2}\right)^2 \left(r+ \frac{m+q}{2}\right)^2}, \\ \label{RN2}
e^{2\beta(r)}&=& \frac{\left(r+\frac{m-q}{2}\right)^2 \left(r+ \frac{m+q}{2}\right)^2}{r^4},
\end{eqnarray}
where $\frac{1}{2}\sqrt{m^{2}-q^{2}} < r< \infty$, and the inversion spherical surface is at $r_{H+}=\frac{1}{2}\sqrt{m^{2}-q^{2}}$, which is also the outer event horizon of the RN spacetime in isotropic coordinates.

The Reisnner-Nordstr\"{o}m metric in isotropic coordinates was first derived in the form given in equations (\ref{RN1}) and (\ref{RN2}) above using the ADM technique (see \citep{Arnowitt1960} and \citep{Arnowitt1960a}).

\textbf{Case 3:} Schwarzschild (S) metric in isotropic coordinates

By setting the parameters $Q= C=\alpha_0=\beta_0=0$, it implies that $\lambda^{2}=1, \phi_{0}=(G)^{-1}, B= \frac{m}{2}= \frac{GM}{2c^{2}}, p_{+}^{2}=1$ and $ p_{-}^{2}=0$ where $m= \frac{GM}{c^{2}}$ is the mass in gravitational units. The metric functions reduce to the well known Schwarzschild solution in isotropic coordinates:
\begin{eqnarray}
e^{2\alpha(r)}= \left( \frac{1- \frac{m}{2r}}{1+ \frac{m}{2r}} \right)^2	\:, \: e^{2\beta(r)}= \left(1+ \frac{m}{2r} \right)^4,
\end{eqnarray}
where $\frac{1}{2} m< r< \infty$ and the inversion spherical surface is at $r_{H}=\frac{1}{2}m$, which is also the event horizon of the Schwarzschild spacetime in isotropic coordinates.

\section{Electrostatics and the Hadamard Method} \label{EP}
We now consider the electrostatic potential due to a ``small" static electric charge $-\epsilon_0 (\vert \epsilon_{0} \vert \ll m_{B})$ situated outside the spherical surface of inversion $B$. 

We let $V(r,\theta,\phi)$ denote the linearly perturbated electrostatic potential so that the perturbed Faraday tensor $F_{ab}$ takes the form
\begin{eqnarray} \label{max}
F_{0i}=-F_{i0}= -c \partial_i V(r,\theta,\phi)\\\nonumber
 F_{ij}=0 \: ; \: i,j=1,2,3 \cdots.
\end{eqnarray}

The perturbed Maxwell equations $\nabla_{[a}F_{bc]}=0$ is automatically satisfied by equation ({\ref{max}). 

Without loss of generality, the perturbed Maxwell equations due to a single electrostatic charge yields
\begin{eqnarray}
\frac{1}{\sqrt{-g}} \partial_b (\sqrt{-g}F^{ab})=J^0
\end{eqnarray}
which implies
\begin{eqnarray} \label{maxwell1.5}
\nabla^2 V(r,\theta, \phi) - (\alpha'(r) - \beta'(r)) \frac{\partial V(r,\theta, \phi)}{\partial r}\\=ce^{2(\alpha(r) +  \beta(r))}J^0
\end{eqnarray}
where the current density $J^0=-\frac{4 \pi \epsilon_0}{cr^2} e^{-2\alpha(r)-3 \beta} \delta(r-b) \delta(\cos \theta- \cos \theta_0)$. Here $\alpha(r)$ and $\beta(r)$ are given by equations (\ref{e2alpha}) and (\ref{e2beta}) respectively, $\nabla^2=\frac{\partial^2}{\partial x^2}+\frac{\partial^2}{\partial y^2}+\frac{\partial^2}{\partial z^2}$ is the 3-dimensional Euclidean space Laplacian with $x=r\sin \theta \cos \phi$, $y=r \sin \theta \sin\phi$ and $z=r\cos \theta$. Note that $\partial_r= \frac{x}{r} \partial_x +\frac{y}{r} \partial_y+\frac{z}{r} \partial_z$. We define
\begin{eqnarray} \label{Gamma} 
\Gamma(r, \theta)=r^2+b^2-2br\cos \theta
\end{eqnarray}
which is equal to the square of the ``radial" distance from the charged particle at $z=b$. 

A brief overview of Hadamard's theory of fundamental solutions \citep{hadamard1923} is neccessary to fully understand Copson's construction (\cite{Copson1928}, \cite{Copson1978}). We adapt Hadamard's result that includes equation (\ref{maxwell1.5}) as a particular case as follows:

\textbf{Theorem 2 (Hadamard's Theorem):}
\noindent \textit{Consider a second order linear partial differential equation of the form 
\begin{eqnarray}  \label{hadamardspde}
\Im(u) =  \sum_{i,j =1}^{3} \delta^{ij} \frac{\partial^2 u}{\partial x^i\partial x^j} + \sum_{i=1}^3 h(r) \frac{x^i}{r} \frac{\partial u}{\partial x^i}  =0 
\end{eqnarray}
where $\delta^{ij}$ is the Kronecker tensor, $h(r)$ is a differentaible function of $r=\delta_{ij} x^{i}x^{j}$. The fundamental solution of equation (\ref{hadamardspde}) is continuous and differentiable everywhere except at the singular point $(r,\theta, \phi)=(b, \theta_0, \phi_0)$ and can be written as 
\begin{eqnarray}
u=   \frac{U(r,\theta,\phi)}{\Gamma^ \frac{1}{2}} 
\end{eqnarray}
where $\Gamma$ is given by equation (\ref{Gamma}) and the function $U(r,\theta, \phi)$ is real analytic everywhere in the domain of definition of equation (\ref{hadamardspde}) including the singular point $(r,\theta, \phi)=(b, \theta_0, \phi_0)$ . $U(r)$ is expandable as a convergent power series in $\Gamma$ such that
\begin{eqnarray} \label{gam}
U(r, \theta, \phi) =   U_0(r) + U_1(r) \Gamma + U_2(r) \Gamma^2 + \cdots,
\end{eqnarray}
where $U_n$ is given by the reccurent formula
\begin{eqnarray} \label{Un}
U_n(r)&=& \frac{U_0}{4 (n-\frac{1}{2}) s} \int^s_0 \frac{s^{n-1}}{U_0} \Im(U_{n-1})ds \\ \nonumber 
 n&=&1,2,3 \cdots \\
s &=& \sqrt{r^2+b^2-2rb \cos \theta} \nonumber
\end{eqnarray}
and 
\begin{eqnarray} \label{U0}
U_0(r)= \exp \left( - \int^r_b h(r) d r \right).
\end{eqnarray}}

In the case of the BDRN metric, the coefficient $h(r)$ in equations (\ref{hadamardspde}) and (\ref{U0}), is given by
\begin{eqnarray}
h(r) = - \alpha'(r) + \beta'(r).
\end{eqnarray}

\section{Extension of the Copson Construction} \label{exten}

Equation (\ref{maxwell1.5}) for the BDRN metric can be expressed in the form
\begin{eqnarray} \nonumber
   \nabla^2 V+ \frac{2B}{r^2-B^2} \left(    2k \left[ \frac{\eta^\ast(r)}{\eta(r)} \right] + \frac{B}{r} \right) \frac{\partial V}{\partial r} \\ \label{maxwell2}
= ce^{2(\alpha(r) +  \beta(r))}J^0,
\end{eqnarray}
where
\begin{eqnarray} \label{k}
k= \frac{C+2}{2\lambda} \quad ; \quad  \eta(r)= p_{+}^{2}-p_{-}^{2}\left(\frac{r-B}{r+B} \right)^{2k},\\
\eta^\ast(r) =  - p_{+}^{2}-p_{-}^{2}\left(\frac{r-B}{r+B} \right)^{2k}.
\end{eqnarray}

Instead of using the formal expression in equation (\ref{Un}) we follow Copson \citep{Copson1928} by substituting equation (\ref{gam}) into (\ref{maxwell2}). After some algebra we obtain the first four terms of the recurrent series of the Brans-Dicke-Reissner-Nordstr\"{o}m metric
\begin{eqnarray}
U_0(r) &=& \frac{r}{b} \frac{\eta_0}{\eta(r)} \frac{(r-B)^{k - \frac{1}{2}}} {(r+B)^{k + \frac{1}{2}}} \frac{(b+B)^{k + \frac{1}{2}}} {(b-B)^{k - \frac{1}{2}}}, \\
U_1 (r)&=& \frac{3B^2 \left( 1+\frac{4}{3}(1-k^2) \right)}{2(r^2-B^2)(b^2-B^2)}U_0,\\
U_2 (r)&=& \frac{B^2  
\left( -5+\frac{4}{3}(1-k^2)\right)}{4(r^2-B^2)(b^2-B^2)}U_1,\\
U_3 (r)&=& \frac{B^2  \left( -7+\frac{4}{3}(1-k^2)  \right)}{10(r^2-B^2)(b^2-B^2)}U_2,
\end{eqnarray}
where $\eta_0=\eta(b)$.
See Table \ref{tab:tab} for the three degenerate cases. 

We introduce the method by which the substitution can be determined for any background with a line element of the form given by equation (\ref{lineelement}) which satisfies the Brans-Dicke electrovac field equations (see Appendix A for the governing equations).
Like Copson, from the first few terms given above we find that the ratio of the ($n+1$)th term to the $n$th term of the power series (\ref{Un}) is proportional to 
\begin{eqnarray}
\frac{B^2}{b^2-B^2}\frac{\Gamma}{r^2-B^2},
\end{eqnarray}
where $r^2-B^2$ is proportional to $\phi r^2 e^{\alpha+ \beta}$ (see equation (\ref{key}) in Appendix A).

Furthermore, the first term of the infinite series, $\frac{U_0}{\Gamma^{1/2}}$, given by equation (\ref{U0}) is proportional to
\begin{eqnarray}
\frac{e^{\frac{1}{2}( \alpha - \beta)}}{\Gamma^{1/2}}= \frac{1}{\gamma^{1/2} \sqrt{\phi} r e^{\beta}}
\end{eqnarray}
where
\begin{eqnarray} \label{potential}
\gamma(r, \theta)=  \frac{B^2}{b^2-B^2} \frac{\Gamma(r,\theta)}{\phi r^{2} e^{\alpha(r)+\beta(r)}}.
\end{eqnarray}

Now we introduce a new dependent variable $F(\gamma)$ such that the perturbed electrostatic potential takes the form
\begin{eqnarray} \label{epsub}
V(r, \theta, \phi)= \frac{F(\gamma)}{\sqrt{\phi} r e^{\beta(r)}}.
\end{eqnarray}

For the Brans-Dicke Reissner-Nordstr\"{o}m background, equation (\ref{epsub}) and (\ref{potential}) become, respectively,
\begin{eqnarray} \label{elec pot}
V(r,\theta)&=& \frac{r \phi_{0}}{\eta(r) (r+B)^{2}} \left( \frac{r-B}{r+B} \right)^{k-1} F(\gamma)\\ \label{gamma}
\gamma(r, \theta)&=&  \frac{B^2}{b^2-B^2}\frac{\Gamma(r, \theta)}{r^{2} - B^{2}}.
\end{eqnarray}

Substituting equations (\ref{elec pot}) and (\ref{gamma}) into (\ref{maxwell2}) gives us a second order linear differential equation on $F(\gamma)$. 
\begin{eqnarray} \nonumber
\gamma(\gamma+1)F''(\gamma)+ \frac{3}{2}(2\gamma +1) F'(\gamma) + (1-k^2)F(\gamma) = 0.  \\  \label{fdiff} 
\end{eqnarray}
We have allowed the right-hand side of the above equation to vanish as we are only interested in regions away from the point cource where the right-hand side of equation (\ref{fdiff}) is zero. We later use our boundary condition to verify that the delta-function source term is satisfied and to also determine the constants of integration of the solution to equation (\ref{fdiff}).

.

\section{Fundamental Solutions} \label{solving}

Equation (\ref{fdiff}) can be solved if we transform the independent variable $\gamma$ as
\begin{eqnarray} \label{subs}
\gamma &=& \sinh^{2} \frac{\zeta}{2},
\end{eqnarray}
which implies 
\begin{eqnarray}
\gamma+1 = \cosh^{2} \frac{\zeta}{2},
\end{eqnarray}
and we write the dependent variable $F(\gamma)$ as
\begin{eqnarray}
F(\gamma)&=& \Phi(\zeta). \label{subs2}
\end{eqnarray}

Using equations (\ref{subs}) and (\ref{subs2}), equation (\ref{fdiff}) can be written in terms of the new variables as follows 
\begin{eqnarray}
\Phi''(\zeta) + 2 \coth \zeta \: \Phi'(\zeta) + (1-k^2) \Phi(\zeta) = 0
\end{eqnarray}
which has the closed-form solution (see \cite{kamke1971})
\begin{eqnarray} 
\Phi(\zeta)= \frac{k}{\sinh \zeta} (\hat{W}_{1}e^{k\zeta}-\hat{W}_{2}e^{-k\zeta})
\end{eqnarray}
where $\hat{W}_{1}$ and $\hat{W}_{2}$ are integration constants. The solution in terms of $\gamma$ is therefore
\begin{eqnarray} \nonumber
F(\gamma)&=& \frac{k}{2\sqrt{\gamma}\sqrt{\gamma+1}} [\hat{W}_{1}(\sqrt{\gamma+1}+\sqrt{\gamma})^{2k} \\
&& \: \: \: - \hat{W}_{2}(\sqrt{\gamma+1}-\sqrt{\gamma})^{2k}]. \label{fgamma}
\end{eqnarray}

Substituting equation (\ref{fgamma}) into equation (\ref{elec pot}) gives the electrostatic potential $V(r,\theta)$ as follows
\begin{eqnarray} \label{solution} \nonumber
V(r,\theta)&=& \frac{k}{2 \eta(r) \sqrt{\gamma}\sqrt{\gamma+1}} \frac{r}{(r+B)^{2}} \left( \frac{r-B}{r+B} \right)^{k-1}  \\
&& \times [\hat{W}_{1}(\sqrt{\gamma+1}+\sqrt{\gamma})^{2k} \nonumber \\ 
&& \: \: \: \: - \hat{W}_{2}(\sqrt{\gamma+1}-\sqrt{\gamma})^{2k} ].
\end{eqnarray}

Consider the inversion point of the static electric point charge $(0,0,(b*))$, where $(b*)= \frac{B^2}{b}$. Let
\begin{eqnarray}
\gamma*= \frac{B^2}{B^2-(b*)^2} \frac{\Gamma*(r,\theta)}{B^2-r^2}\\
\Gamma*= r^2 +(b*)^2 -2(b*)r \cos \theta.
\end{eqnarray}
Thus $\Gamma*$ is equal to the square of the ``radial" distance from the inversion point at $z=(b*)$. It is straightforward to verify that
\begin{eqnarray}
\gamma + 1 = \gamma*.
\end{eqnarray}

The electrostatic potential $V(r,\theta)$ in equation (\ref{solution}) is therefore singular at the point charge $z=b$ and also at its inversion point $z=(b*)$. One can also verify that as the field point $r$ approaches the inversion surface $r=B$, the potential approaches a finite limit value provided that $C>-2$.

Finally, to determine the fundamental solution for the electrostatic potential, which allows only one free parameter to arise from the presence of the perturbing electrostatic charge, it is necessary to establish the relationship between the two arbritrary constants in equation (\ref{solution}).

\section{Determination of Integration Constants} \label{surfintandbc}

In 1927, Whittaker, using the method of separation of variables in the usual Schwarzschild coordinates, found the solution expressing the electrostatic potential of a charge in a Schwarzschild background as an infinite series \citep{Whittaker1927b}. His result was later confirmed by Cohen and Wald \citep{cohen_point_1971} in 1971 and Hanni and Ruffini \citep{hanni_lines_1973} in 1973. A commonality of these works is the use of a boundary condition stating that a charge should not arise inside the horizon as a result of the presence of the perturbing electric charge situated outside the horizon.

This boundary condition was not implemented by Copson in his determination of integration constants in \citep{Copson1928} and \citep{Copson1978} due to the fact that the region inside the horizon is excised in the isotropic coordinates. Instead Copson chose values for the integration constants such that the overall solution would be symmetric in interchanging the position of the field point $r$ with the position  of the perturbing charge $b$. As a result his solution, as he pointed out himself, was in contradiction to Whittaker's solution by the existence of a non-vanishing zero-order term. Linet \citep{Linet1976}, using the boundary condition at infinity and Gauss' theorem, found this second charge which was necessarily excised to give a result which was in accordance to those given by \citep{Whittaker1927b}, \citep{hanni_lines_1973}, and \citep{cohen_point_1971}.

In \citep{Linet1979}, Linet and Teyssandier found a single closed-form solution describing the electrostatic potential generated by a perturbing charge in a Brans-Dicke background. They expressed the fundamental solution as a sum of this solution and legendre functions before performing a multipole expansion and writing the fundamental solution completely in terms of legendre functions. By expressing the solution as a multipole expansion they are then able to impose boundary conditions at infinity to get a meaningful solution upon which the Gauss' theorem can then be implemented to yield their final closed-form solution.

Here we introduce a method of determining the integration constants of equation (\ref{solution}) which does not require one to expand the closed-form solution into an infinite series and which is even more stringent than those set by \citep{Whittaker1927b}, \citep{cohen_point_1971} and \citep{Hanni1973}. We impose the condition that any integration over a closed spatial region not containing the perturbing charge must be exactly zero even if that area contains a surface of inversion. Naturally, an integration over an area containing the perturbing charged particle must therefore equal exactly $4 \pi \epsilon$, where $\epsilon$ is the charge of the particle. From Appendix B we know that for the Brans-Dicke Reissner-Nordstr\"{o}m background, the generalized Gauss's theorem can be written as the following

\begin{widetext}
\begin{eqnarray} \label{int}
\int_{\Re} J^{0} d \upsilon =\int_{0}^{2\pi}  \int_{-\pi}^{\pi}  \eta(r)^{2} (r+B)^{2}  \left(\frac{r-B}{r+B} \right)^{\frac{\lambda - C - 2}{\lambda}} 
 \frac{\partial V(r)}{\partial r} \sin \theta d\theta d\phi. 
\end{eqnarray}
\end{widetext}

Here, $\Re$ is a region of 3-dimensional space residing in a hypersurface and $\partial \Re$ is its closed 2-dimensional boundary. Again, $d \upsilon$ is an element of spatial proper volume in $\Re$. 
In order to integrate the above we convert equation (\ref{solution}) into a function of $\sinh \zeta$ where $\gamma = \sinh^{2} \frac{\zeta}{2}$. We find that the only term that requires integration is the term containing the integration constants, the integral of which is 
\begin{eqnarray} \nonumber
\int_{-\pi}^\pi \frac{\hat{W}_1 e^{k\zeta} - \hat{W}_2 e^{-k\zeta}}{\sinh \zeta} \sin \theta d \theta =\\ \nonumber \frac{2brB^2[(b+B)^{2k} - (b-B)^{2k}]}{k (b^2-B^2)^{k-1}(r^2-b^2)^{k-1}} \\ \label{no charge} \times[\hat{W}_1 (r+B)^{2k} - \hat{W}_2 (r-B)^{2k}] 
\end{eqnarray}
for $B<r<b$ and
\begin{eqnarray}\nonumber
\int_{-\pi}^\pi \frac{\hat{W}_1 e^{k\zeta} - \hat{W}_2 e^{-k\zeta}}{\sinh \zeta} \sin \theta d \theta =\\ \nonumber \frac{2brB^2[(r+B)^{2k} - (r-B)^{2k}]}{k (b^2-B^2)^{k-1}(r^2-b^2)^{k-1}} \\ \label{charge} \times[\hat{W}_1 (b+B)^{2k} - \hat{W}_2 (b-B)^{2k}] 
\end{eqnarray}
for $B<b<r$.

When we return equation (\ref{no charge}) into equation (\ref{int}) it is fairly straightforward to see that for the electrostatic potential to vanish for the region not containing a charge the integration constants must be chosen as the following
\begin{eqnarray}
\hat{W}_1= p^2_+ \hat{W}\\
\hat{W}_2=p^2_- \hat{W},
\end{eqnarray}
where $\hat{W}$ is a constant yet to be determined. 

By returning equation (\ref{charge}) into equation (\ref{int}) and under the condition that for this region ($B<b<r$) the equation (\ref{int}) must equal $-4 \pi \epsilon_0$ we can quickly solve for $\hat{W}$ giving 
\begin{eqnarray}
\hat{W} = \frac{bB (b^2-B^2)^{k-1} \sqrt{\phi_0} \epsilon_0}{k [p_+^2 (b+B)^{2k} - p_-^2 (b-B)^{2k}]}.
\end{eqnarray}

\section{Degenerate Cases} \label{specificcases}

The relationship between the four cases and the process by which one reduces to the other is made obvious in Table \ref{tab:tab}. It is straightforward to convert the equations in the Reissner-Nordstr\"{o}m and Schwarzschild spacetimes into their more familiar form when one conducts the transformations given in Section \ref{STFT} of this paper. When the transformations are made we find that the solutions are in agreement with the closed-form solutions given by Linet \cite{Linet1976}, Leaute and Linet  \citep{leaute_electrostatics_1976} and Linet and Teyssandier \citep{Linet1979} and with multipole expansions given by Hanni and Ruffini \citep{hanni_lines_1973} and Cohen and Wald \citep{cohen_point_1971}.

\section{Conclusion}
In this paper we have established an ansatz to solve the perturbed Maxwell equations due to an electrostatic charge in a Brans-type spacetime in isotropic coordinates which is reducible to the Schwarzschild and Reissner-Nordstr\"{o}m black hole solutions by extending Copson's method. As Copson's solution is based on Hadamard's theory of fundamental solutions of general Laplace equations it would be interesting to see whether Hadamard's infinite series converges to give Copson's closed-form result.

By finding the coefficients to $U_0, U_1, U_2, \cdots$ through the direct substitution of  Hadamard's infinite series into the field equations one is able to compare them with the coefficients given in this paper using Copson's method. 

In a seperate paper, a formal proof of Hadamard's fundamental solution equation (\ref{hadamardspde}) is given. We find that Copson's results in \citep{Copson1928} are in fact exactly equal to those found using Hadamard's method and go on to investigate how the Hadamard method relates to the results obtained by Linet in \citep{Linet2005}. We also find that the discrepancy between Copson and Hadamard with those from the literature including Whittaker, Hanni and Ruffini and Cohen and Wald lies in the domain of definition of fundamental solutions in the presence of a surface of inversion when considering the situation in isotropic coordinates.

Furthermore, we investigate the scope of applicability of Hadamard's theorem including its application to more general Scalar-Tensor-Vector theories and $f(R)$ theory and in particular to other branches of the Brans-Dicke theory such as the Barker and Schwinger cases (see also \citep{Singh1984}). For detailed discussions on the scope of applicability of the Copson method in higher dimensions see \citep{Linet2005}.

In a separate paper, we convert the results given in this paper from isotropic coordinates to the usual Schwarzschild coordinates using Linet's transformation (outlined in \citep{Linet1976}) and plot equipotential surfaces in both coordinate systems. As alluded to in Item \ref{bhadra} of Section \ref{STFT}, \citep{Bhadra2005} found that Brans-Dicke black holes cannot exist if the weak-field approximation is to be upheld. In our next paper we postulate that the weak-field approximation need not be satisfied \citep{Barcelo2000}, \citep{McInnes2002}, \citep{Wu1986}, \citep{Hawking1975}. Thus we find it worthwhile to plot the results of this paper in the usual coordinates to gain better insight into the behavior of the scalar field inside the horizon and thereby shed light on the physical possibility of Brans-Dicke black holes.

\section*{Acknowledgements}
One of the authors of this paper (M.W) would like to thank the Australian Government for the International Postgraduate Research Scholarship and Monash University for the Monash Graduate Scholarship.

The authors would also like to thank S. Deser for his helpful advice and an anonymous referee for their useful suggestions.

\begin{sidewaystable*}
\vspace*{8cm} 
\hspace*{-2.5cm}
\begin{tabular}{|c|c|c|c|c|}
\hline 
  & Brans-Dicke-Reissner-Nordstr\"{o}m & Brans-Dicke & Reissner-Nordstr\"{o}m & Schwarzschild \\ 
\hline

{ \rule{0cm}{0.7cm} $e^{2\alpha}$  } 
& {$\frac{e^{2\alpha_0} \left|\frac{r-B}{r+B} \right|^{\frac{2}{\lambda}}}{\eta(r)^{2}}$  } 
& { $e^{2\alpha_0}\left|\frac{r-B}{r+B} \right|^{\frac{2}{\lambda}}$ } 
& { $\frac{\left(\frac{r-B}{r+B} \right)^2}{\eta(r)^{2}}$ }  
& { $\left(\frac{r-B}{r+B} \right)^2$  }  \\ 
\hline

{ \rule{0cm}{0.7cm} $e^{2\beta}$ } 
& {$  e^{2\beta_0} \eta(r)^2  \left( 1 + \frac{B}{r} \right)^{4} \left| \frac{r-B}{r+B} \right|^{2 \left( \frac{\lambda-C-1}{\lambda} \right)} $ }  
& {$ e^{2\beta_0} \left( 1 + \frac{B}{r} \right)^{4} \left| \frac{r-B}{r+B} \right|^{2 \left( \frac{\lambda-C-1}{\lambda} \right)}$ } 
& {$ \eta(r)^2 \left( 1 + \frac{B}{r} \right)^{4} $ } 
& {$\left( 1 + \frac{B}{r} \right)^{4}$ }  \\ 
\hline

{ \rule{0cm}{0.7cm} $\eta(r)$  } 
& { $ p_{+}^{2}-p_{-}^{2}\left|\frac{r-B}{r+B} \right|^{2k}$ } 
& { $1$  }  
& {$ p_{+}^{2}-p_{-}^{2}\left(\frac{r-B}{r+B} \right)^{2}$ } 
& {$1 $ }  \\ 
\hline

{ \rule{0cm}{0.7cm} $\phi$ } 
& {$\phi_{0} \left| \frac{r-B}{r+B} \right|^{\frac{C}{\lambda}}$   } 
& {$\phi_{0} \left| \frac{r-B}{r+B} \right|^{\frac{C}{\lambda}}$  } 
& { $\phi_0$  } 
& { $\phi_0$ }  \\
\hline

{\rule{0cm}{0.7cm} $U_{0}$   } 
& {$\frac{r}{b} \frac{\eta_0}{\eta(r)} \frac{(r-B)^{k - \frac{1}{2}}} {(r+B)^{k + \frac{1}{2}}} \frac{(b+B)^{k + \frac{1}{2}}} {(b-B)^{k - \frac{1}{2}}}$  }
& { $\frac{r}{b} \frac{(r-B)^{k - \frac{1}{2}}}{(r+B)^{k + \frac{1}{2}}} \frac{(b+B)^{k + \frac{1}{2}}}{(b-B)^{k - \frac{1}{2}}} $  } 
& { $\frac{r}{b} \frac{\eta_0}{\eta(r)} \frac{(r-B)^{\frac{1}{2}}}{(r+B)^{\frac{3}{2}}} \frac{(b+B)^{\frac{3}{2}}}{(b-B)^{\frac{1}{2}}} $  } 
& { $\frac{r}{b}  \frac{(r-B)^{\frac{1}{2}}}{(r+B)^{\frac{3}{2}}} \frac{(b+B)^{\frac{3}{2}}}{(b-B)^{\frac{1}{2}}} $  }  \\
\hline 

{\rule{0cm}{0.7cm} $U_{1}$   } 
& { $\frac{3B^2 \left( 1+\frac{4}{3}(1-k^2) \right)}{2(r^2-B^2)(b^2-B^2)}U_0$ } 
& { $\frac{3B^2 \left( 1+\frac{4}{3}(1-k^2) \right)}{2(r^2-B^2)(b^2-B^2)}U_0$  } 
& { $\frac{3B^2}{2(r^2-B^2)(b^2-B^2)}U_0$   } 
& { $\frac{3B^2}{2(r^2-B^2)(b^2-B^2)}U_0$   }  \\
\hline 

{\rule{0cm}{0.7cm} $U_{2}$   } 
& {$\frac{B^2 \left( 1+\frac{4}{3}(1-k^2) \right) 
\left( -5+\frac{4}{3}(1-k^2)\right)}{8(r^2-B^2)^2(b^2-B^2)^2}U_0$   } 
&{$\frac{B^2 \left( 1+\frac{4}{3}(1-k^2) \right) 
\left( -5+\frac{4}{3}(1-k^2)\right)}{8(r^2-B^2)^2(b^2-B^2)^2}U_0$   } 
& { $-\frac{5B^4}{8(r^2-B^2)^2(b^2-B^2)^2}U_0$ } 
& { $-\frac{5B^4}{8(r^2-B^2)^2(b^2-B^2)^2}U_0$  }  \\
\hline 

{\rule{0cm}{0.7cm} $U_{3}$  } 
& {$\frac{B^6 \left( 1+\frac{4}{3}(1-k^2)  \right) \left( -5+\frac{4}{3}(1-k^2) \right) \left( -7+\frac{4}{3}(1-k^2)  \right)}{80(r^2-B^2)^3(b^2-B^2)^3}U_0$  } 
& { $\frac{B^6 \left( 1+\frac{4}{3}(1-k^2)  \right) \left( -5+\frac{4}{3}(1-k^2) \right) \left( -7+\frac{4}{3}(1-k^2)  \right)}{80(r^2-B^2)^3(b^2-B^2)^3}U_0$ } 
& {$-\frac{7B^6}{80(r^2-B^2)^3(b^2-B^2)^3}U_0$  } 
& {$-\frac{7B^6}{80(r^2-B^2)^3(b^2-B^2)^3}U_0$  }  \\
\hline 

{\rule{0cm}{0.7cm} Eq. (\ref{epsub})  }  
& { $\frac{r \phi_{0}}{\eta(r)(r+B)^{2}} \left( \frac{r-B}{r+B} \right)^{k-1} F(\gamma)$   } 
& {$\frac{r}{(r+B)^{2}} \left( \frac{r-B}{r+B} \right)^{k-1} F(\gamma)$   } 
& { $\frac{r \phi_{0}}{\eta(r)(r+B)^{2}}  F(\gamma)$   } 
& { $\frac{r \phi_0}{(r+B)^{2}} F(\gamma)$ }  \\ 
\hline

{\rule{0cm}{0.7cm} $V(r,\theta)$ }

& \parbox[t]{6cm}{$\frac{\epsilon_0 r}{\eta(r) (r^2-B^2)} \left[ \frac{r-B}{r+B} \right]^{k}  \frac{bB (b^2-B^2)^{k-1}}{[p_+^2(b+B)^{2k}-p_-^2(b-B)^{2k}]} $ 
  \\  $\times \frac{p_+^2(\sqrt{\gamma+ 1}+ \sqrt{\gamma})^{2k} - p_-^2(\sqrt{\gamma+ 1}- \sqrt{\gamma})^{2k}  }{2\sqrt{\gamma}\sqrt{\gamma+1}}$ }

&  \parbox[t]{5cm}{$  \frac{\epsilon_0 r}{ r^2-B^2} \left[ \frac{r-B}{r+B} \right]^{k}  \frac{bB (b^2-B^2)^{k-1}}{(b+B)^{2k}}$
 \\ 
 $ \times\frac{(\sqrt{\gamma+ 1}+ \sqrt{\gamma})^{2k} }{2\sqrt{\gamma}\sqrt{\gamma+1}}$ }
  
&  \parbox[t]{5cm}{$ \frac{\epsilon_0 r}{\eta(r) (r+B)^2}   \frac{bB }{[p_+^2(b+B)^{2}-p_-^2(b-B)^{2}]}$
 \\ 
 $ \times \frac{p_+^2(\sqrt{\gamma+ 1}+ \sqrt{\gamma})^{2} - p_-^2(\sqrt{\gamma+ 1}- \sqrt{\gamma})^{2}  }{2\sqrt{\gamma}\sqrt{\gamma+1}}$ 
}  
& \parbox[t]{2cm}{$ \frac{\epsilon_0 r}{ (r+B)^2}   \frac{bB }{(b+B)^{2}} $
 \\ 
$ \times  \frac{(\sqrt{\gamma+ 1}+ \sqrt{\gamma})^{2} }{2\sqrt{\gamma}\sqrt{\gamma+1}}$ 
 }\\
\hline 
\end{tabular}
\caption{The four cases and their solutions for the electrostatic potential generated by a charged particle at $r=b$ where $k=\frac{C+2}{2 \lambda}$}
\label{tab:tab}
\end{sidewaystable*} 
\cleardoublepage
\pagebreak
\newpage

\section*{Appendix A: Brans-Dicke Reissner-Nordstr\"{o}m Background} \label{appendixa}

The Brans-Dicke electrovac equations (\ref{fieldone}), (\ref{fieldtwo}) and (\ref{maxwell}) arising from the static spherically symmetric line element (\ref{lineelement}) in isotropic coordinates can be simplified when the following substitutions are introduced 
\begin{eqnarray} \label{tildeA}
\tilde{A}(r) &:=& \alpha(r) + \frac{1}{2}[\ln \phi(r)],\\ \label{tildeB}
\tilde{B}(r) &:=& \beta(r) + \frac{1}{2}[\ln \phi(r)].
\end{eqnarray}
The electrovac equations ($e^{2\beta}G^t_t$, $e^{2\beta}G^r_r$ and $e^{2\beta}(G^r_r +G^\theta_\theta)$) from equation (\ref{fieldone}) can then be written as
\begin{eqnarray} \label{gtt}
  2 \tilde{B}''(r) + \tilde{B}'(r)^2 + \frac{4}{r} \tilde{B}'(r) + \frac{4 \pi Q^2 e^{-2\tilde{B}(r)}}{c^4 } \rule{1.5cm}{0cm} \nonumber \\ + \frac{2\omega+3}{4} ([\ln \phi(r)]')^2 =0, \rule{0.5cm}{0cm}\\ \label{grr}
  \tilde{B}'(r)^2 + 2\tilde{A}'(r) \tilde{B}'(r) + \frac{2}{r} (\tilde{A}'(r) + \tilde{B}'(r)) 
\rule{1.8cm}{0cm} \nonumber \\ + \frac{4 \pi Q^2 e^{-2\tilde{B}(r)}}{c^4}  - \frac{2\omega+3}{4} ([\ln \phi(r)]')^2= 0, \rule{0.5cm}{0cm}  \\
\label{grrp}
   \tilde{A}''(r) + \tilde{B}''(r) + ((\tilde{A}(r)+ \tilde{B})')^2  \rule{3cm}{0cm} \nonumber \\ + \frac{3}{r} (\tilde{A}'(r) +\tilde{B}'(r))= 0, \rule{0.5cm}{0cm} 
\end{eqnarray}

The above three equations are not linearly independent, but instead are related via the following
\begin{eqnarray}
- \tilde{A}'(r) e^{2\beta} G^t_t + \left( \frac{d}{dr}+ \left(\tilde{A}'(r)+ 2\tilde{B}'(r) + \frac{4}{r} \right) \right) e^{2\beta}G^r_r \nonumber \\  - 2 \left(\tilde{B}'(r) + \frac{1}{r} \right) e^{2\beta} (G^r_r + G^\theta_\theta) =0. \rule{1cm}{0cm}
\end{eqnarray}

We point out here that the integrations below are carried out formally without taking ito account the signature or actual boundary values of $\tilde{A}_b$, $\tilde{B}_b$, $\tilde{A}_{b}'+\tilde{B}_{b}'$, $\phi_b$ and $\phi_{b}'$ where the former are the corresponding values of $\tilde{A}(r)$, $\tilde{B}(r)$, $\tilde{A}(r)'+\tilde{B}(r)'$, $\phi(r)$ and $\phi(r)'$ at the boundary point at infinity. 

Equation (\ref{grrp}) can be expressed as a Cauchy-Euler equation
\begin{eqnarray} \label{grr2}
(e^{\tilde{A}(r)+\tilde{B}(r)})'' + \frac{3}{r} (e^{\tilde{A}(r)+\tilde{B}(r)})' =0,
\end{eqnarray}
which can be solved to give 
\begin{eqnarray} \label{key}
e^{\tilde{A}(r)+\tilde{B}(r)} = e^{\tilde{A}_b+\tilde{B}_b} \left(1- \frac{\varepsilon^2 B^2}{r^2} \right),
\end{eqnarray}
and 
\begin{eqnarray}
\lim_{r \rightarrow \infty}r^3 (\tilde{A}'(r) + \tilde{B}'(r))= 2 \varepsilon^2 B^2,\\
\varepsilon^2 \in \lbrace -1, +1 \rbrace. \rule{1.7cm}{0cm}
\end{eqnarray}

The reduced long-range scalar field wave equation (equation (\ref{scalarfield})) can be written in terms of $\tilde{A}$ and $\tilde{B}$ as
\begin{eqnarray} \label{scalarwave}
\left(  \frac{r^2 e^{\tilde{A}+\tilde{B}}\phi'(r)} {\phi} \right)'=0.
\end{eqnarray}
By integrating equation (\ref{scalarwave}) twice from $r$ to infinity we obtain
\begin{eqnarray} \label{36}
\phi= \phi_0 \left( \frac{r-\varepsilon B}{r+ \varepsilon B} \right)^{\frac{C}{\varepsilon \lambda}},
\end{eqnarray}
where
\begin{equation}
\lambda^2= \frac{\varepsilon^2}{4} \left((2 \omega+3) C^2+(C+2)^2  \right) >0.
\end{equation}

We rewrite the modified field equation (\ref{grr}) into the following form
\begin{eqnarray} \label{40}
\tilde{A}'(r)^2= (\tilde{A}+ \tilde{B})' \left(\tilde{A}' + \tilde{B}' + \frac{2}{r} \right)  \rule{1 cm}{0cm} \nonumber \\ + \frac{4 \pi Q^2 e^{-2\tilde{B}(r)}}{c^4 } - \frac{2\omega+3}{4} ([\ln \phi(r)]')^2 \rule{0.25cm}{0cm}
\end{eqnarray}

Using equations (\ref{key}) and (\ref{36}), after some algebra we obtain a first order second degree separable differential equation
\begin{widetext}
\begin{eqnarray}
\left( \frac{d}{dr} \left( e^{-\tilde{A}(r)} \right)  \right)^2&=& \frac{4 \pi Q^2 e^{-2(\tilde{A}_b + \tilde{B}_b)}}{c^4 (r^2- \varepsilon^2 B^2)^2} \left[ \left( \frac{c^4 B^2 e^{2(\tilde{A}_b + \tilde{B}_b)} (C+2)^2}{4\pi Q^2 \lambda^2}  \right) e^{-2\tilde{A}(r)} + 1 \right].
\end{eqnarray}
Since $e^{\tilde{A}_b}=\sqrt{\phi_0}e^{\alpha_b}$ and $e^{\tilde{B}_b}=\sqrt{\phi_0}e^{\beta_b}$ the solution to this equation gives
\begin{eqnarray} \label{alphafinal}
e^{-\alpha(r)} =  e^{-\alpha_b} \left( \frac{r-\varepsilon B}{r+ \varepsilon B} \right)^{\frac{C}{2 \varepsilon \lambda}} \left( p_{+}^2 \left( \frac{r-\varepsilon B}{r+ \varepsilon B} \right)^{\frac{C+2}{2 \varepsilon \lambda}} - p_{-}^2  \left( \frac{r-\varepsilon B}{r+ \varepsilon B} \right)^{-\frac{C}{2 \varepsilon \lambda}} \right),
\end{eqnarray}
where $p_+$ and $p_-$ are given in equation (\ref{pnp}).
\begin{eqnarray} \label{betafinal}
e^{\beta(r)} = e^{\beta_b} \left( 1+ \frac{\varepsilon B^2}{r^2} \right) \left( \frac{r-\varepsilon B}{r+ \varepsilon B} \right)^{-\frac{C}{2 \varepsilon \lambda}} \left( p_{+}^2 \left( \frac{r-\varepsilon B}{r+ \varepsilon B} \right)^{\frac{C+2}{2 \varepsilon \lambda}} - p_{-}^2  \left( \frac{r-\varepsilon B}{r+ \varepsilon B} \right)^{-\frac{C}{2 \varepsilon \lambda}} \right).
\end{eqnarray}
\end{widetext}
When $\varepsilon^2=+1$ the above coincides with the BDRN metric given in Theorem 1 in Section (\ref{STFT}) above. The solutions corresponding to the Brans Type II, and Type III and IV solutions are given by setting, respectively, $\epsilon^2=-1$ and taking the limit when $\epsilon\rightarrow 0$  (using L'Hopital's rule) on equations (\ref{36}), (\ref{alphafinal}) and (\ref{betafinal}).

\section*{Appendix B:  Gauss' Theorem} \label{appendixc}

In order to determine the integration constants in equation (\ref{solution}) we use Gauss's theorem; a brief overview of which is given here.
Let $\Re$ be a region of 3-dimensional space residing in a hypersurface $\varrho$ and let $\partial \Re$ be its closed 2-dimensional boundary. Gauss' theorem states that for the electric field $E^a$ (and indeed for any given vector field, see Wald \citep{Wald})
\begin{eqnarray}
\int_{\Re} \nabla_a E^a d\upsilon = \int_{\partial \Re} E^a \cdot n_a dS \label{surfint}
\end{eqnarray}
where $d\upsilon$ is an element of spatial proper volume in $\Re$, $n_a$ is the outward facing unit vector orthogonal to the closed 2-dimensional boundary $\partial \Re$ and $dS$ is the usual surface element $dS=r^{2}\sin \theta d\theta d\phi$.

We know that the electric field is related to the Faraday tensor by the following 
\begin{eqnarray}
E^a = F^{ab} n_b.
\end{eqnarray}  
Using the above and equation (\ref{max}) we find that the electric field is indeed equal to the gradient of the electrostatic potential $V(r,\theta,\phi)$ and therefore the right-hand side of equation (\ref{surfint}) can be written as 
\begin{eqnarray}
\int_{d\Re} E^a \cdot n_a dS = \int_{d\Re} \nabla V \cdot \hat{n} dS.
\end{eqnarray}

From Maxwell's equations the left-hand side of equation (\ref{surfint}) can be written as $\int_{\Re} J^{0} d \upsilon$ where 
\begin{equation}
J^0=-\frac{4 \pi \epsilon_0}{cr^2} e^{-2\alpha(r)-3 \beta} \delta(r-b) \delta(\cos \theta- \cos \theta_0)
\end{equation}
is the charge density. 

It follows that for the region $B<b<r$ containing the point charge $-\epsilon_{0}$ positioned at $r=b, \theta=0$ the left-hand side of equation (\ref{surfint}) becomes $-4 \pi \epsilon_{0}$ and for any region not containing the charge, i.e. $B<r<b$, the left-hand side vanishes. 

For the Brans-Dicke Reisnner-Nordstr\"{o}m spacetime as $n^a$ is orthogonal to $\partial \Re$ the only term that remains is the $r$ term and equation (\ref{surfint}) becomes
\begin{eqnarray} \nonumber
\int_{\Re} J^{0} d \upsilon = \eta(r)^{2} (r+B)^{2} \left(\frac{r-B}{r+B} \right)^{1-2k}   \\ \times \int_{0}^{2\pi}  \int_{-\pi}^{\pi}  \frac{\partial V(r, \theta)}{\partial r} \sin \theta d\theta d\phi. \label{intre}
\end{eqnarray}

$\rule{0cm}{1cm}$ The left-hand side of equation ({\ref{intre}) is determined by whether or not the region $\Re$ contains the singular point at $r=b$. In particular for the purposes of this investigation, the theorem determines the choice of integration constants in $V(r, \theta)$ as can be seen in the main section of this article.

\bibliography{forarXiv}

\end{document}